\documentclass[prd,preprint,nofootinbib,tightenlines,showpacs,amsmath,amssymb,floatfix]{revtex4}

\def\be{\begin{equation}}
\def\ee{\end{equation}}
\def\bea{\begin{eqnarray}}
\def\eea{\end{eqnarray}}

\begin{document}

\title{Spacetime Averaged Null Energy Condition}

\author{Douglas Urban}
\email{Douglas.Urban@tufts.edu}
\author{Ken D. Olum}
\email{kdo@cosmos.phy.tufts.edu}
\affiliation{Institute of
  Cosmology, Department of Physics and Astronomy, Tufts University,
  Medford, MA 02155, USA.}


\begin{abstract}
The averaged null energy condition has known violations for quantum
fields in curved space, even when one considers only achronal geodesics.
Many such examples involve rapid variation in the stress-energy tensor
in the vicinity of the geodesic under consideration, giving rise to
the possibility that averaging in additional dimensions would yield a
principle universally obeyed by quantum fields.  However, after
discussing various procedures for additional averaging, including
integrating over all dimensions of the manifold, we give here a class of
examples that violate any such averaged condition.
\end{abstract}

\maketitle

\section{Introduction}
General relativity itself places no restrictions on which geometries
may be considered. If we want to exclude pathological phenomena, such
as closed timelike curves, wormholes, or superluminal communication,
we must appeal to restrictions on the stress-energy tensor of the
sources, $T_{ab}$.  For classical fields there is a variety of such
{\em energy conditions} which all hold and have been used to exclude
such behavior
\cite{Friedman:1993ty,Olum:1998mu,Tipler:1976bi,Hawking:1991nk}.

The null energy condition (NEC) states that for a null vector $l^a$
we always have $T_{ab}l^al^b \geq 0$. Other pointlike conditions often
discussed, such as the weak, strong and dominant conditions, all imply
the null energy condition. However, all pointlike conditions are
violated by quantum fields. Even a simple vacuum plus two photon state
possesses negative energy density in some regions.

We can instead consider the averaged null energy condition (ANEC),
\be\label{eqn:ANEC}
\int_\gamma d\lambda\, T_{ab}l^al^b \geq 0
\ee
where $\gamma$ is a null geodesic, $l^a$ the tangent to it, and
$\lambda$ an affine parameter.  Equation~(\ref{eqn:ANEC}) can easily be
violated with compactified dimensions, but we can restrict our
attention to the case where $\gamma$ is an achronal geodesic,
i.e., where no two points of $\gamma$ are also connected by a timelike
path.  Positive energy densities lead to gravitational lensing, which
produces conjugate points, so in the absence of negative energy densities
generic geodesics are chronal. 

Requiring ANEC to hold only on achronal geodesics eliminates many
known violations, while still ruling out many exotic phenomena
\cite{Graham:2007va}.  ANEC always holds in Minkowski space
\cite{Klinkhammer:1991ki,Wald:1991xn} (where all null geodesics are
achronal), and along achronal geodesics surrounded by a flat tubular
neighborhood \cite{Fewster:2006uf} in curved space.  Nevertheless, even
``achronal ANEC'' can be violated.  One kind of violation found by Visser \cite{Visser:1994jb} can be
produced by the scale anomaly for conformally coupled scalar fields. This effect is logarithmically
suppressed and depends on the renormalization scale.

In addition, we recently found two ANEC violations that are present in
conformally flat spacetimes (where Visser's anomaly vanishes) and do
not depend on the renormalization scale \cite{Urban:2009yt}.
Conformal transformations of conformally coupled fields are an easy
case to analyze, as the transformation properties of the stress tensor
are known \cite{Page:1982fm}. If we write the conformally flat metric
$\bar{g}_{ab}=\Omega^2\eta_{ab}$, the conformally coupled scalar field
transforms as $\bar{\phi}=\Omega^{-1}\phi$, but the stress tensor does
not transform simply.  Instead, it has anomalous terms, so
\be
\bar{T}_{ab}=\Omega^{-2}T_{ab}+\text{curvature anomaly}.
\ee

Our first construction utilizes a sequence of states in
Minkowski space whose momentum is constrained to lie within a cone
that lengthens and narrows as a parameter $\alpha \to 0$. As the limit
is taken the states approach the vacuum, but the stress tensor near the
origin grows more negative. Although the Minkowski space ANEC integral
vanishes in the $\alpha \to 0$ limit, in a transformed space the
integrand is weighted by $\Omega^{-4}$, which can enhance the negative
contribution. For small enough $\alpha$ the curvature anomaly can be
dwarfed by this negative contribution.

The second violation uses only the curvature anomaly, which can
itself be negative. One can consider transforming the ground state, so
that the $\Omega^{-2}T_{ab}$ contribution to $\bar{T}_{ab}$
vanishes. It is worth noting that the transformation of the ground
state is not in general the ground state in the new space. Using null
coordinates defined in Minkowski space, with $u=(z-t)/\sqrt{2}$ and
$v=(z+t)/\sqrt{2}$, the specific example illustrated in
Ref.~\cite{Urban:2009yt} is
\be \label{Wold}
\Omega={\rm exp}\left[\left(a+\frac{bx^2}{r^2}\right)e^{-(u^2+v^2+x^2+y^2)/r^2}   \right]
\ee

In both of these situations, the magnitude of the violation grows as
the effect is more tightly constrained to the geodesic.  This
motivates us to inquire whether a version of ANEC that includes
averaging in additional directions would avoid these violations.
Indeed, if one's average includes timelike directions, one can use the
null-contracted, timelike-averaged quantum energy inequality of
Ref.~\cite{FewsterRomanwitherratum} to limit the amount by which such
an additionally averaged ANEC can be violated.  In such a case, it is
not possible to produce a sequence of states that give unlimited
violation, as we did in Ref.~\cite{Urban:2009yt} for ANEC alone.

For the curvature case, Eq.~(\ref{Wold}) gives the ANEC integral
\be
16\sqrt{2\pi}a(b-2a)\beta/r^3\,,
\ee
where
\be\label{eqn:beta}
\beta=-\frac{1}{5760\pi^2}\,.
\ee
Since $\beta<0$, if we choose $b > 2a$, ANEC will be violated.  In
this particular case, additional averaging gives a positive result.
Nevertheless, we show below that similar constructions can violate any
generally averaged condition, and thus no additionally averaged
version of ANEC is generally obeyed in curved spacetimes.

We work in units where $c=1$ and $\hbar = 1$.  Our sign conventions
are (+++) in the categorization of Misner, Thorne, and Wheeler
\cite{mtw}.

\section{Additionally Averaged Null Energy Conditions}

Averaging the null energy over a null geodesic eliminates many
violations of the NEC, but even ANEC has violations. It might be
that averaging in additional directions could eliminate these and
yield a principle that all quantum fields would obey.  But what do we
mean by a more general average of NEC?  If we establish a null vector
field $l^a$ throughout spacetime, we can project the stress-energy
tensor on this field and take the average,
\be\label{4ave}
A_4=\int \sqrt{-g}d^4x T_{ab}l^a l^b \,,
\ee
but is not clear how we should define $l^a$.

In the case of the regular ANEC, we can start with a vector $l^a$
tangent to our null geodesic $\gamma$ at some initial point $p$.  Such
a vector is defined only up to rescaling, but such change (equivalent
to a change of affine parameter) only affects the magnitude of the
ANEC integral, not its sign.  We then establish $l^a$ everywhere on
the geodesic by parallel transport from $p$ to each destination point
$x$.

We could attempt the same technique for averaging in more dimensions,
but now there is more than one choice of path for the parallel
transport.  In general, when we work in curved space the resulting
$l^a$ will depend on the path chosen.  Flanagan and
Wald~\cite{Flanagan:1996gw} make the choice to transport $l^a$ along
a geodesic from $p$ to $x$.  This is well defined if one works inside
a normal neighborhood.  If one considers perturbations of flat space
as done in Ref.~\cite{Flanagan:1996gw}, and as we will do below, one
can transport $l^a$ in the unperturbed space-time without ambiguity.
But in the general case, there may be no geodesic, or multiple
geodesics, connecting $p$ and $x$, and the procedure does not work.

We can also consider averaging over more than a single geodesic but
less than all the dimensions of the manifold.  For example, let $\chi$
be a timelike line parametrized by proper time $\tau$.  Start with a
null vector $l^a$ at some point $p\in\chi$, and establish a null
vector field $l^a$  on $\chi$ by parallel transport.  Through each
point of $\chi$ draw the null geodesic whose tangent vector is $l^a$.  Then
we can write
\be\label{2ave}
A_2=\int d\tau d\lambda T_{ab}l^a l^b 
\ee

Similarly, we can average only over spacelike directions, but here we
will encounter ambiguities.  Given a spacelike 2-surface $\Sigma$, let
us establish a null vector field $l^a$ orthogonal to the surface at
each point.  These vectors generate a family of geodesics.  We can
take the integral over each one, to get
\be\label{3ave1}
A_3=\int_\Sigma \sqrt{g_2}\,d\sigma_1 d\sigma_2 \int T_{ab}l^a l^b d\lambda\,.
\ee
Here $\sigma_1$ and $\sigma_2$ are the coordinates on the surface and $g_2$ the
induced metric.  The inner integral is to be taken over the geodesic
generated by $l_a$ at each point.

The direction of $l^a$ is fixed by orthogonality, but we need to fix
the magnitude.  As before, we could try do to this via parallel
transport, but that may depend on the path chosen.  Thus
this idea does not lead to a well-defined averaging procedure.

This process depends on the choice of the initial surface, even if the
resulting null 3-surface is fixed.  Suppose we propagate our initial
surface an affine distance $\lambda$ down each geodesic to get a new
surface $\Sigma'$.  The geodesics may spread out or squeeze together
between $\Sigma$ and $\Sigma'$.  Thus if we started with $\Sigma'$
instead of $\Sigma$, we would have a different weighting of the
geodesics.  To avoid this problem, we could integrate over the surface
for each $\lambda$ first and then combine them, giving
\be\label{3ave2}
A'_3=\int d\lambda \int_{\Sigma(\lambda)} \sqrt{g_2}\,d\sigma_1 d\sigma_2
T_{ab}l^a l^b\,.
\ee
However, Eq.~(\ref{3ave2}), like Eq.~(\ref{4ave}), is not in an obvious
way an average of ANEC.

We will not attempt to solve these problems, but rather we will
exhibit counterexamples that apply to a very wide class of averaging
procedures.  We are able to do this because we work to first
nonvanishing order in a spacetime that is a small perturbation of flat
space.  As we did in Ref.~\cite{Urban:2009yt}, we work in a
conformally flat spacetime with conformal factor $\Omega = e^\omega
\sim 1+\omega$, with $\omega\ll 1$.  We define our average by letting
$l^a$ be constant in the unperturbed spacetime and find violations of
averaged versions of ANEC at order $\omega^2$.  Suppose now that we
use a different procedure.  If we defined $l^a$ by parallel transport
along a path which winds many times in the region where $\omega$ is
largest, we could of course accumulate a large change in $l^a$.  But
this procedure is obviously pathological.  If we restrict ourselves to
a path which is free of such windings, the change in $l^a$ along a
path $C$ will be given schematically by
\be
\Delta l^a\sim\int_C \Gamma^a_{bc} l^b dC^c
\ee
If the scale of the curved region is given by $r$, the magnitude of 
$\Gamma^a_{bc}$ is of order $\omega/r$, so $\Delta l^a\sim \omega$.
Thus the effect of the choice of path is of higher order in $\omega$
than the original effect and can be consistently neglected.

\section{Curvature Anomaly}
We will generate violations using the anomalous curvature term in the
conformal transformation of $T_{ab}$. First we will review the
curvature anomaly for a general conformally flat space, before
specifying a transformation. A conformally coupled field transforms as
$\bar{\phi}=\Omega^{-1}\phi$, but the stress tensor has extra terms,
given fully in \cite{Page:1982fm}. Our analysis follows
\cite{Urban:2009yt}. When beginning with Minkowski space, the Weyl
tensor always vanishes. Terms proportional to $g_{ab}$ vanish upon
null projection, while those proportional to $R_{;ab}$ vanish on
integration along each geodesic.  Considering only the remaining
terms, we have
\be
\bar{T}_{ab}=\Omega^{-2}T_{ab} +
2\beta\left[\bar{R}^c_a\bar{R}_{cb} -
  \bar{R}\bar{R}_{ab} \right]\,,
\ee
where $\beta$ is given by Eq.~(\ref{eqn:beta}).

Next we express these curvature quantities in terms of the conformal transformation, with $\omega=\ln\Omega$.  Again dropping terms with $g_{ab}$, the stress tensor is given by
\bea
\bar{T}_{ab}=\Omega^{-2}T_{ab} + 8\beta\Omega^{-2} \big[ {\omega^{,c}}_a\omega_{,cb} -2
  \left(\Box\omega+\omega^{,c}\omega_{,c}\right)\left(\omega_{,ab}-\omega_{,a}\omega_{,b}\right) \nonumber \\
   - \omega^{,c}\omega_{,a}\omega_{,cb} - \omega^{,c}\omega_{,b}\omega_{,ca} \big]
\eea
We take an initial state with $T_{ab}=0$, so the state does not
contribute to $\bar T_{ab}$. We also have $\omega$ much less than
one, so we may ignore terms of order $\omega^3$ and take
$\Omega\approx 1$. That leaves us with only
\be
\bar{T}_{vv}=8\beta\big[g^{cd}\omega_{,cv}\omega_{,dv} -2
 \Box\omega\omega_{,vv}  \big]
\ee
In our coordinates we organize this as
\be
\bar{T}_{vv}=8\beta\big[\omega_{,xv}^2 + \omega_{,yv}^2 -2
 \left(\omega_{,uv}+\omega_{,xx}+\omega_{,yy}\right)\omega_{,vv}  \big]
\ee

We first study a particular transformation which gives a simpler
violation of ANEC than \eqref{Wold}, which is
\be
\omega=axr^{-1}e^{-\rho}\,,
\ee
where we define $\rho=(u^2+v^2+x^2+y^2)/r^2$. This gives a localized transformation, so our spacetime is both conformally and asymptotically flat. The stress tensor component at $x=y=0$ is
\be
\bar{T}_{vv}=\frac{32a^2v^2\beta}{r^6}e^{-2\rho}
\ee
Because $\beta$ is a negative number, this is always negative.  Thus
integrating over $\gamma$ for any fixed $u$ violates ANEC, so of
course $A_2$ is negative as well. This gives a violation of greater
magnitude as $a$ grows, but this analysis depends on $a\ll 1$ so it is
not possible to build an arbitrarily large violation.

For averaging transversely and over the whole spacetime the above example does not give a negative answer, so instead we use
\be
\omega=\left(bu+cv\right)r^{-1}e^{-\rho}
\ee
Any term which is odd in $x,\,y,$ or $z$ will vanish on integration,
so we do not write such terms.  Including only the even terms, the
$vv$ stress tensor component becomes
\bea \label{fullT}
\bar{T}_{vv}&=&\frac{32\beta}{r^{10}}\bigg\{4b^2u^2\left[(x^2+y^2)(r^2-v^2)+r^2(2v^2-r^2)\right]
   \nonumber \\
&&-2bc\left[8u^2v^4-2r^2(v^4+5u^2v^2) +r^4(u^2+3v^2) \right]\nonumber \\
&&+c^2\left[(x^2+y^2)(-4v^4+8v^2r^2+r^4)+4r^2(2v^4-3r^2v^2)\right]\bigg\}e^{-2\rho}
\eea
To calculate $A_3$ (which coincides with $A_3'$ to first order) we set $u=0$, and this becomes
\be
\bar{T}_{vv}=\frac{32\beta}{r^{10}}\left\{2bc(2v^4r^2-3v^2r^4)
+c^2\left[(x^2+y^2)(-4v^4+8v^2r^2+r^4)+4r^2(2v^4-3r^2v^2)\right]\right\}e^{-2\rho}.
\ee
Note that here, the $b^2$ term drops completely, but the $c^2$ term is entirely unchanged. The integral is
\be
A_3= -\frac{3\sqrt{2}\pi^{3/2}\beta}{r} c(2b+c)
\ee
So long as $b<-c/2$ this will be a negative quantity.  

If instead we average \eqref{fullT} over the whole manifold, as in \eqref{4ave}, we have
\be
A_4 =-\beta \pi^2\left( b^2+6bc+3c^2  \right)
\ee
For $-3-\sqrt{6} < b/c < -3+\sqrt{6}$, the average is negative. As
discussed earlier, these results still hold even if the integrand
differs by any power of $\Omega$. Here all dependence on $r$ has
dropped, and thus the sharpness of the curvature does not affect the
violation.

\section{Discussion}
We have demonstrated counterexamples for generally averaged null
energy conditions, with averaging over a timelike 2-surface, a null
3-surface, and the entire manifold.  We now find it unlikely there is
any principle that would rule out exotic phenomena by restricting the
total set of possible stress-energy tensor configurations without
regard to the background.  If so, we must restrict our attention to
self-consistent fields, that is, quantum fields that give rise to an
exotic spacetime with that same spacetime, rather than an arbitrary
different one, as the background.  In other words, the entire system
should satisfy the semiclassical Einstein equation,
\be
G_{ab}=8\pi G \left(\langle T^{\text{quantum}}_{ab}\rangle + T_{ab}^{\text{classical}}\right)\,.
\ee
Here $T^{\text{classical}}$ is a classical stress tensor that obeys
energy conditions, and $T^{\text{quantum}}$ is the stress tensor of
some state of a set of quantum fields, all in the background whose Einstein
tensor is $G_{ab}$.

In the previous sections, we found violations of ANEC and its averages
in small perturbations around Minkowski space.  The same situation was
studied for self-consistent systems by Flanagan and Wald
\cite{Flanagan:1996gw}, who found that ANEC was obeyed, at least for
pure states.  In fact, no violations of ``self-consistent achronal
ANEC'' are known, and it is powerful enough to rule out many
problematic spacetimes \cite{Graham:2007va}.  Nevertheless, it could
be the case that even this condition needs to be weakened by
additional averaging.  It is also possible that a negative yet finite state-independent lower bound gives a useful condition, as discussed in Ref. \cite{Yurtsever:1995gf}. The first example considered in Ref. \cite{Urban:2009yt} violates singly averaged ANEC without bound, but with additional averaging the examples considered here are all finite. Unfortunately, it is difficult to define such
an average outside of perturbation theory, as discussed in Sec.~II.

\section*{Acknowledgments}

The authors thank Chris Fewster and Tom Roman for helpful discussions.
This research was supported in part by Grants No. RFP1-06-024 and No.
RFP2-06-23 from The Foundational Questions Institute (fqxi.org).

\end{document}